\documentclass[12pt]{article}
\usepackage{amsmath}
\usepackage{amsfonts}
\usepackage{amsthm}
\normalbaselines
\catcode `\@=11
\@addtoreset{equation}{section}

\def\section{\@startsection {section}{1}{\z@}{-3.5ex plus -1ex minus
     -.2ex}{2.3ex plus .2ex}{\normalsize\bf}}
\def\subsection{\@startsection{subsection}{2}{\z@}{-3.25ex plus -1ex minus
 -.2ex}{1.5ex plus .2ex}{\normalsize\bf}}

\def\thebibliography#1{\section*{References\markboth
  {REFERENCES}{REFERENCES}}\list
  {[\arabic{enumi}]}{\settowidth\labelwidth{[#1]}\leftmargin\labelwidth
  \advance\leftmargin\labelsep
  \usecounter{enumi}}
  \def\newblock{\hskip .11em plus .33em minus -.07em}
  \sloppy
  \sfcode`\.=1000\relax}
 
\catcode `\@=12
\def\<#1|#2>{\left\langle #1\vphantom{#2}\right|\left.\negmedspace%
\vphantom{#1}#2\right\rangle}
\newcommand{\cA}{{\cal A}}
\newcommand{\cE}{{\cal E}}
\newcommand{\cL}{{\cal L}}
\newcommand{\cK}{{\cal K}}
\newcommand{\C}{\mathbb{C}}
\newcommand{\Q}{\mathbb{Q}}
\newcommand{\R}{\mathbb{R}}
\newcommand{\Z}{\mathbb{Z}}
\newcommand\CS{C$^*$}
\newcommand{\emf}[1]{{\bfseries\boldmath #1}} 
\DeclareMathOperator{\spec}{spec}
\DeclareMathOperator{\tr}{tr}

\begin{document}

\vspace*{2.5cm}
\noindent
{ \bf NON-COMMUTATIVE BLOCH THEORY: AN OVERVIEW}\vspace{1.3cm}\\
\noindent
\hspace*{1in}
%
\begin{minipage}{13cm}
Michael J.\ Gruber\vspace{0.3cm}\\
Institute of Mathematics, Humboldt University at Berlin, Germany \\
E-mail: 
 {gruber@mathematik.hu-berlin.de} \\
WWW: 
 {http://spectrum.mathematik.hu-berlin.de/\~{}gruber}
\end{minipage}

\vspace*{0.5cm}

\begin{abstract}
\noindent
For differential operators which are invariant under the 
action of an abelian group Bloch theory is the tool
of choice to analyze spectral properties. By shedding some 
new non-commutative light on this we motivate
the introduction of a non-commutative Bloch theory for elliptic 
operators on Hilbert C$^*$-modules. It relates
properties of C$^*$-algebras to spectral properties of module
operators such as band structure, weak genericity
of cantor spectra, and absence of discrete spectrum. It 
applies e.g.\ to differential operators invariant under 
a projective group action, such as Schr\"odinger operators
with periodic magnetic field.
\end{abstract}


\section*{INTRODUCTION}
Bloch (or Floquet) theory in its usual form already has a long history. 
Basically it starts from the fact that partial differential equations with 
constant coefficients are mapped into algebraic equations by means of the 
Fourier or Laplace transform. 
Now, if the coefficients are not constant but just periodic under an abelian 
(locally compact topological) group one still has the Fourier transform on 
such groups, mapping functions on the group $\Gamma$ into functions on the 
dual group $\hat\Gamma$; 
the original spectral problem on a non-compact manifold is mapped into a 
(continuous) sum of spectral problems on a compact manifold (see section 
\ref{CBT}).

This is what makes Bloch theory an indispensible tool especially for solid 
state physics, where one describes the motion of non-interacting electrons 
in a periodic solid crystal by a Schr\"odinger operator
$ -\Delta + V$ on $L^2(\R^d)$. 
The potential function $V$ is the gross electric potential generated by all 
the crystal ions and thus is periodic under the lattice given by the crystal 
symmetry.

Measurements of crystals often require magnetic fields $b$ (2-form). 
In quantum mechanics, they are described by a vector potential (1-form) $a$ 
such that $b=da$ ($B=\operatorname{curl}A$ for the corresponding vector fields). 
The magnetic Schr\"odinger operator 
then reads \[ H=-(\nabla-ia)^2+V. \]

But, although $b$ is periodic or even constant, $a$ need not be so, and $H$ 
won't be periodic. 
It is therefore necessary to use magnetic translations (first introduced by 
Zak \cite{Zak:DESEF}) under which $H$ still is invariant. 
But now, these translations do not commute with each other in general. 
Therefore ordinary (commutative) Bloch theory does not apply. 

Basically, the reason for this failure is that a non-abelian group has no 
``good'' group dual: the set of (equivalence classes of) irreducible 
representations has no natural group structure whereas the set of 
one-dimensional representations is too small to describe the group --- 
otherwise it would be abelian.

But although $\hat\Gamma$ does not exist any more, the algebra $C(\hat\Gamma)$
 of continuous functions continues to exist in some sense:
It is given by the reduced group \CS-algebra of $\Gamma$ which is just the 
\CS-algeba generated by $\Gamma$ in its regular representation on itself 
(on $l^2(\Gamma)$).

Section \ref{CBTNPV} shows how one can re-formulate ordinary Bloch theory in 
a way which refrains from using the points of $\hat\Gamma$ and relies just on
 the r\^ole of $C(\hat\Gamma)$.
{}From a technical point of view this requires switching from measurable fields
 of Hilbert spaces to continuous fields which then can be described as Hilbert 
\CS-modules over the commutative \CS-algebra $C(\hat\Gamma)$.

Having done this one can retain the setup but omit the condition of 
commutativity for the \CS-algebra $C(\hat\Gamma)$.
Thus one is lead to non-commutative Bloch theory (section \ref{NBT}) dealing
 with elliptic operators on Hilbert \CS-modules over non-commutative 
\CS-algebras.
The basic task is now to relate properties of the \CS-algebra to spectral 
properties of ``periodic'' operators.
Thus one generalizes spectral results for elliptic operators on compact 
manifolds as well as  results of ordinary Bloch theory.

In section \ref{E} we list examples where non-commutative Bloch theory applies.

\bigskip
This article is an overview of a part of my Ph.D.\ thesis \cite{Gru:NB} which 
is written in German.
Due to space limitations the following sections will be rather sketchy. 
A full account of that part in English is in preparation \cite{Gru:NBT}, as 
well as for the other, related parts \cite{Gru:BTQMS,Gru:MFSASCSMSO}. 
I am indepted to my thesis advisor Jochen Br\"uning for scientific support.

\bigskip

This work has been supported financially by Deutsche Forschungsgemeinschaft 
(DFG) as project D6 at the SFB 288 (differential geometry and quantum 
physics), Berlin.

\section{\hspace{-4mm}.\hspace{2mm}COMMUTATIVE BLOCH THEORY}\label{CBT}
\subsection*{
Setup}
Let $X$ be a smooth oriented Riemannian manifold and
$\Gamma$ a discrete abelian group, acting on $X$
properly discontinuously ($\Rightarrow M:=X/\Gamma$ is Hausdorff), freely 
($\Rightarrow M$ smooth), isometrically ($\Rightarrow M$ Riemannian), and 
co-compactly ($\Leftrightarrow M$ compact). 
The $\Gamma$-action on $X$ induces an action on $C^\infty_{(c)}(X)$ and a 
unitary action on $L^2(X)$ via
\[ (\gamma_*f)(x) := f(\gamma^{-1}x) \]
for $x\in X,\gamma\in\Gamma$ and $f$ in the corresponding space of functions.

Let
$D$ be a symmetric $\Gamma$-periodic elliptic differential operator on $X$, 
i.e.\ on its domain of definition $C_c^\infty(X)\subset L^2(X)$. By 
\emf{$\Gamma$-periodic} we mean that $D$ commutes with the $\Gamma$-action on its 
domain.

The basic physical example is  $X=\R^d$ ($d=2,3$), $\Gamma=\Z^d$ acting by 
translations (or magnetic translations) and $D$ given by the Schr\"odinger 
operator (or magnetic Schr\"odinger operator with integral flux) with 
periodic electric potential.

\subsection*{
Aim}
Our aim is to determine the type (set/measure theoretic) of the spectrum of $D$. 
By set theoretic type of 
the spectrum\footnote{Under the aforementioned conditions $D$ is essentially 
self-adjoint, so that the closure of $D$ is the only self-adjoint extension 
and has only real spectrum. By abuse of notation we denote the closure by 
$D$, too. } 
we mean either \emf{band structure} (i.e.\ a locally finite union of closed intervals)
 or \emf{Cantor 
structure} (i.e.\ a nowhere dense set without isolated points). 
Bands may degenerate to points which would not be called bands by physicists. 
Non-degenerate bands allow formation of (semi-) conductors.

Measure theoretic properties of the spectrum are continuity properties of the 
spectral measure with respect to Lebesgue measure. 
Physically one expects either pure point spectrum (eigenvalues and their 
accumulation points) or absolutely continuous spectrum (bands).
Thus one wants to exclude the third possibility: singular continuous spectrum.

\subsection*{
Method}
The basic and well-known method for the spectral theory of periodic elliptic 
operators is \emf{Bloch theory}, which in one dimension is also called 
\emf{Floquet theory}. 
Its first step is to construct a direct integral 
\begin{align} L^2(X)&\simeq\int^\oplus_{\hat\Gamma}H_\chi\,d\chi,\\
D&\simeq\int^\oplus_{\hat\Gamma}D_\chi\,d\chi, \\ 
\intertext{ where the fiber Hilbert spaces }
H_\chi&=L^2(F_\chi) \\\intertext{ are spaces of square-integrable sections of 
associated complex line bundles}
F_\chi&=X\times_\chi\C,\\\intertext{ and the operators in the fiber are given
 by the gauge-periodic boundary conditions} 
D_\chi&=D|_{C^\infty(F_\chi)}. 
\end{align}
The decomposition
$\Phi:L^2(X)\rightarrow \int_{\hat\Gamma}^\oplus H_\chi\,d\chi$ is defined by
\begin{align} 
(\Phi f)(x)_\chi := \sum_{\gamma\in\Gamma}\chi(\gamma)f(\gamma^{-1}x)
\end{align}
for $f\in C^\infty_c(X),\chi\in\hat\Gamma,x\in X$ and can be extended 
unitarily to $L^2(X)$.

$\hat\Gamma$ may be identified with the Brillouin zone in solid state physics,
 $H_\chi$ is the space of wave functions with quasi-momentum $\chi$. 
The family $(H_\chi)_{\chi\in\hat\Gamma}$ is a measurable field of Hilbert 
spaces; decomposability of $D$ w.r.t.\ this field is equivalent to 
$\Gamma$-periodicity of $D$.

The decomposition described above is still valid for the magnetic 
Schr\"odinger operator with zero magnetic flux per lattice cell but has to 
be modified for non-zero integral flux. 
In any case, the fibers $D_\chi$ may be identified with magnetic Schr\"odinger
 operators on the quotient space $M=X/\Gamma$ on which there may be 
inequivalent quantizations of the classical magnetic system.
Indeed, the family $(D_\chi)_{\chi\in\hat\Gamma}$ contains all possible 
quantization classes (\cite{AscOveSei:MBABL,Gru:BTQMS}).

\subsection*{
Results}
By general results for direct integrals (see e.g.\ \cite{Dix:AOEHAN}, chapter 
II, \S 1) one can compute the spectrum of $D$ from the spectra of the family 
$(D_\chi)_{\chi\in\hat\Gamma}$.
Using special properties of this family one gets:
\begin{enumerate}
\item
Since $(D_\chi)_{\chi\in\hat\Gamma}$ is a continuous family
 of operators with compact resolvent, the spectrum of $D$ is given as the union
 $\spec D=\bigcup\limits_{\chi\in\hat\Gamma}\spec D_\chi$ and thus
has band structure. Bands may degenerate to points, but possible eigenvalues 
have infinite multiplicity automatically.
\item
 Using the real analyticity of the operator family one gets:
\begin{itemize}
\item $\spec_{s.c.}D=\emptyset$
\item $\spec_{p.p.}D$ is discrete as a subset of $\R$.
\end{itemize}
For the magnetic Schr\"odinger operator with zero magnetic flux this is due
 to \cite{HemHer:BGPMH}; for rational flux (and general abelian-periodic 
elliptic operators) this was done in
\cite{Gru:NB,Gru:MFSASCSMSO}.
\end{enumerate}


\section{\hspace{-4mm}.\hspace{2mm}COMMUTATIVE BLOCH-THEORY FROM A 
NON-COMMUTATIVE POINT OF VIEW}\label{CBTNPV}
As seen above it is necessary to use, in addition to a measurable field of 
Hilbert spaces, the continuity property of an operator family.
Thus the basic idea is to incorporate the continuity into the setup, i.e.¸ to 
find a continuous sub-field.
Now, a continuous field of Hilbert spaces over a space $\hat\Gamma$ is 
equivalent to a Hilbert \CS-module over $C(\hat\Gamma)$.
In our geometric context such a module is given naturally:
For $e,f\in C^\infty_c(X)$ define 
\begin{align} \<e|f>_\cE(\chi):=\<\Phi(e)_\chi|\Phi(f)_\chi>_{H_\chi}. 
\end{align}
 This gives a $C(\hat\Gamma)$-linear pre-scalar product, completion gives a 
Hilbert $C(\hat\Gamma)$-module $\cE$, periodic operators are adjointable 
module operators.

How to get back $L^2(X)$ from $\cE$? This can be done by means of the Hilbert 
GNS representation:
Haar measure $d\chi$ on $\hat\Gamma$ defines a faithful state $\tau$ on 
$C(\hat\Gamma)$ via integration and
\begin{align} 
\<e|f>_\tau=\int_{\hat\Gamma}\<\Phi(e)_\chi|\Phi(f)_\chi>_{H_\chi}=
\<e|f>_{L^2(X)} \end{align}
so that the representation space $\cE_\tau$ is just $L^2(X)$.

The second basic observation is that $C(\hat\Gamma)=C^*_{red}(\Gamma)$ is 
 the reduced group \CS-algebra of $\Gamma$, i.e.\ the \CS-algebra generated 
by $\Gamma$ in its regular representation on $l^2(\Gamma)$. 
This algebra continues to exist for non-abelian groups, but will be 
non-commutative. 

\section{\hspace{-4mm}.\hspace{2mm}NON-COMMUTATIVE BLOCH THEORY}\label{NBT}
\subsection*{
Setup}
Let $\cA$ be a \CS-algebra and $H$ a Hilbert space which is a right 
$\cA$-module.
Let $D$ be a (possibly unbounded) self-adjoint operator on $H$, commuting 
with the module action of $\cA$.
For physical examples we refer to section \ref{E}.

\subsection*{
Aim}
We now want to investigate the relations between $\spec D$ and $\cA$; 
in particular this should reproduce 
the band structure results in the commutative case as described above.

\subsection*{
Method}
The basic step is to construct a Hilbert $\cA$-module $\cE$ and a faithful 
(tracial) state $\tau$ on $\cal A$ such that the Hilbert GNS representation 
gives back the Hilbert space on which to do spectral theory: 
$H\simeq\cE_\tau$; 
and such that $D$ comes from an unbounded self-adjoint module operator $F$ on 
$\cal E$ which is $\cA$-elliptic (see below).
This construction has to be done for each class of examples separately and may
  require hard analytic work; 
once they fit into the general framework it is just (\CS-) algebraic propeties
 which are used.

Under these assumptions one can construct a trace $\tr_\tau$ on the 
$\tau$-trace class $\cL_\cA^1(\cE,\tr_\tau)$ in the module operators which 
generalizes the trace per unit volume in solid state physics.
Applying this trace to projections one gets as usual a generalized dimension 
$\dim_\tau$ for the range of projections.

\subsection*{
Ellipticity}
Let $T$ be an unbounded operator on $\cE$. $T$ is called \emf{$\cA$-elliptic} if
\begin{enumerate}
\item $T$ is densely defined,
\item $T$ is regular, i.e.\ $T^*$ exists, is densely defined, and 
$\operatorname{ran}(1+T^*T)$ is dense in $\cE$, and
\item $T$ has $\cA$-compact resolvent, i.e.\ $(1+T^*T)^{-1}\in\cK_\cA(\cE)$.
\end{enumerate}
This is the notion of ellipticity which is usual for operators on Hilbert modules.


\subsection*{
Basic criteria}
Let $\cal C$ be a \CS-algebra, $\tau$ a trace. $\cal C$ has the 
\emf{Kadison property} if there
 is $c>0$ such that for all non-zero projections $P$ in $\cal C$ one has $\tau(P)\geq c$.
\\
Let $\cal C$ be a \CS-algebra, $\tau$ a state. $\cal C$ has \emf{real rank zero with 
inifinitesimal state} if every self-adjoint element can be approximated by
 a finite spectrum element with arbitrary small $\tau$-value on the spectral
 projections.

\subsection*{
Results}
\begin{enumerate}
\item 
If $\lambda$ is an isolated eigenvalue of $D$ then the corresponding eigenspace 
$H_\lambda$ is 
an (algebraically) finitely generated projective Hilbert $\cA$-module.\\
If $e^{-D^2}$ is of $\tau$-trace class then $H_\lambda$ has finite $\tau$-dimension:
 $\dim_\tau H_\lambda<\infty$\\
If $\cE,\cA$ are ``suitable'' then $H_\lambda$ is infinite dimensional 
($\dim H_\lambda=\infty$),
 in particular the discrete spectrum is empty:
$\spec_{disc}D=\emptyset$

\item 
If $\cK_\cA(\cE)$ has the Kadison property and $e^{-D^2}$ is of $\tau$-trace
 class then
$D$ has band spectrum (the basic idea going back to 
\cite{Sun:GCASPSOM,BruSun:SPEO,BruSun:SGPEO}).

\item 
If $\cK_\cA(\cE)$ has real rank zero with infinitesimal state ($RRI_0$) then
 Cantor spectrum is weakly generic (\cite{ChoEll:DSAEFSIRCA}), i.e.\ every 
operator
can be approximated by ones with Cantor spectrum in norm resolvent sense.
\end{enumerate}

The first part is analogous to the case of elliptic operators on compact 
manifolds: these have compact resolvent and therefore finite-dimensional 
eigenspaces, whereas in our situation we have $\cA$-compact resolvent and 
finitely generated modules, but (under suitable conditions) 
infinite-dimensional eigenspaces.

The second part traces back band structure to a property that holds in the 
commutative case.

The third part gives a citerion for weakly generic (i.e.\ for a dense set of 
operators) total break-down of band structure.

\section{\hspace{-4mm}.\hspace{2mm}EXAMPLES}\label{E}
\subsection*{
Commutative Bloch theory}
$\cA$ is the algebra of continuous functions $C(\hat\Gamma)$ on the 
character group, 
$\cE$ the space of sections of a continuous field of Hilbert spaces 
defined by
the continuous Bloch sections. 
The state $\tau$ is given by integration w.r.t.\ Haar measure: 
$\tau(f)=\int_{\hat\Gamma}f(\chi)\,d\chi$. From this it follows that  
$C(\hat\Gamma)$ has the Kadison 
property which implies band structure.
Furthermore, we are in the ``suitable'' situation so that any possible 
eigenspace is infinite-dimensional but has finite $\tau$-dimension.

\subsection*{
Periodic elliptic operators}
Here $\cA$ is the reduced group \CS-algebra $C^*_{red}(\Gamma)$ of $\Gamma$,
$\cE$ is defined by 
\[ \<e|f>_\cE:=\sum_{\gamma\in\Gamma}\<T_\gamma e|f>_{L^2(E)}L_\gamma \]
 for a vector bundle $E$ over $X$ with lift $T_\gamma$ of the $\Gamma$-action;
 $\tau$ is the canonical trace, $L_\gamma$ the left regular representation of 
$\Gamma$ on $l^2(\Gamma)$.
This reproduces \cite{Sun:GCASPSOM,BruSun:SPEO}.

\subsection*{
Gauge-periodic elliptic operators}
This case is as above, but addtionally with a projective lift $U_\gamma$ of 
the action such that \[ U_
{\gamma_1}U_{\gamma_2}=\Theta(\gamma_1,\gamma_2)U_{\gamma_1\gamma_2}. \]
Therefore $\Theta$ defines a group cohomology class $[\Theta]\in 
H^2(\Gamma,S^1)$, and
$\cA=C^*_{red}(\Gamma,\Theta)$ is a twisted reduced group \CS-algebra,
$\cE\simeq\cA\otimes h$ as above.
In particular $\cA$ is a rotation algebra $\cA_\alpha$ for the $\Z^2$-periodic
 magnetic Schr\"odinger operator, where $\alpha$ denotes the magnetic flux.

If $\alpha\in\Q$ then  $\cA$ has the Kadison property so that 
$\cK_\cA(\cE)$ has the Kadison property, too,  and $D$ has band spectrum 
(reproducing 
\cite{BruSun:SGPEO}).

If $\alpha\in\R\setminus\Q$ then $\cA$ has $RRI_0$ so that 
$\cK_\cA(\cE)$ has $RRI_0$, too, (\cite{EllEva:SIRCA}) and Cantor spectrum is
 weakly generic.

\subsection*{
Hofstadter model, quantum pendulum}
This is the case of the difference equations known as almost Matthieu, 
Hofstadter type or quantum pendulum, arising in several models in solid state 
physics (Peierls substitution, mesoscopic systems) as well as in integrable
 systems.
Here we have just a trivial Hilbert module $\cA=\cE=\cA_\alpha$ over a 
rotation algebra. 
Therefore, the results are as above.


\end{document}